\documentclass[twocolumn,showpacs,preprintnumbers,amsmath,amssymb]{revtex4}

\usepackage{graphicx}
\usepackage{dcolumn}
\usepackage{bm}

\usepackage{wrapfig,rotating}
\usepackage{amssymb,amsmath,array}
\usepackage{pslatex}
\usepackage{hyperref} 

\newcommand{\sqrtsnn}{\sqrt{s_{\mbox{\tiny{\it{NN}}}}}}
\newcommand{\sqrts}{\sqrt{s}}

\newcommand{\jpsi}{J/\psi}

\newcommand{\ups}{\Upsilon}

\newcommand{\pom}{I\!\!P}
\newcommand{\gaga}{\gamma\,\gamma}
\newcommand{\gp}{\gamma\mbox{-p}}
\newcommand{\gA}{\gamma\mbox{-A}}

\providecommand{\qqbar}{Q\overline{Q}}
\providecommand{\elel}{e^+e^-}
\providecommand{\mumu}{\mu^+\mu^-}
\providecommand{\lele}{l^{+}\,l^{-}}

\begin{document}


\title{Forward Physics at the LHC: within and beyond the Standard Model }

\author{David d'Enterria}
\affiliation{CERN, PH, CH-1211 Geneva 23, Switzerland}

\date{\today}

\begin{abstract}
We review the detection capabilities in the forward direction of the various LHC experiments 
together with the associated physics programme. A selection of measurements accessible with 
near-beam instrumentation in various sectors (and extensions) of the Standard Model (SM) is 
outlined, including QCD (diffractive and elastic scattering, low-$x$ parton dynamics, 
hadronic Monte Carlos for cosmic-rays), electroweak processes in $\gaga$ interactions,
and Higgs physics (vector-boson-fusion and central exclusive production). 
\end{abstract}


\maketitle

\section{Introduction}
\label{sec:intro}

\begin{figure}[htbp]
\includegraphics[width=0.45\columnwidth,height=3.3cm]{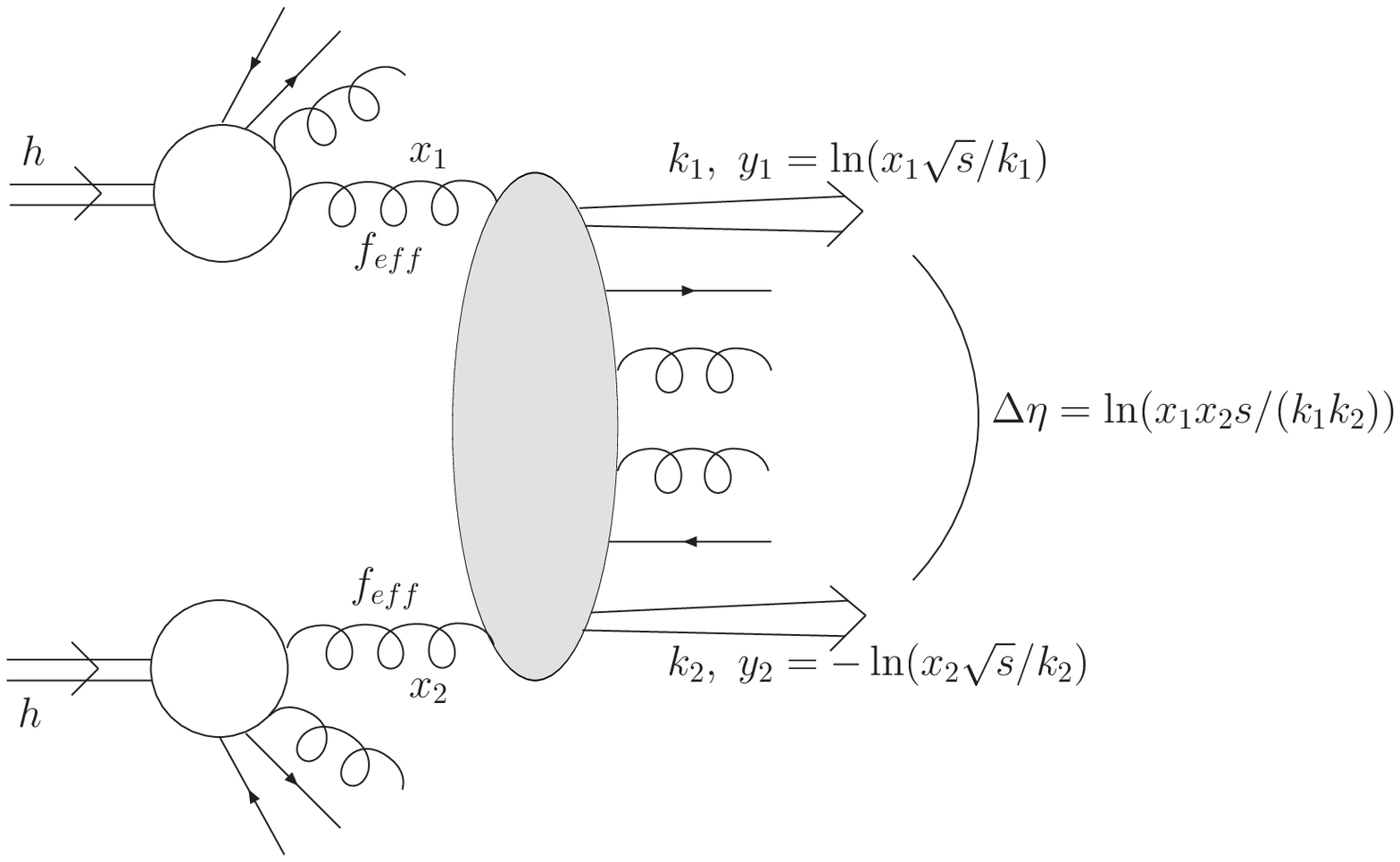}
\includegraphics[width=0.45\columnwidth,height=3.0cm]{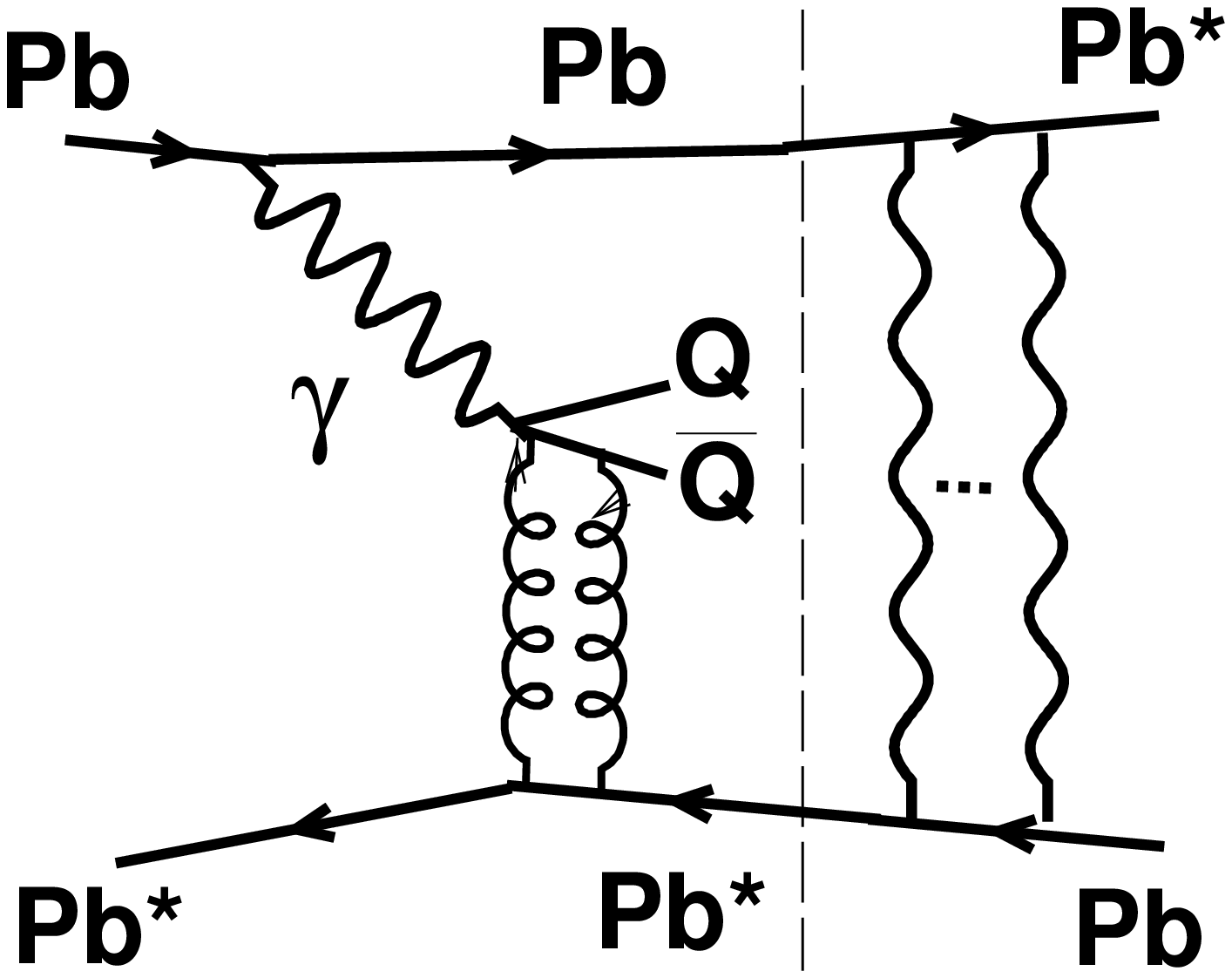}
\includegraphics[width=0.41\columnwidth,height=2.7cm]{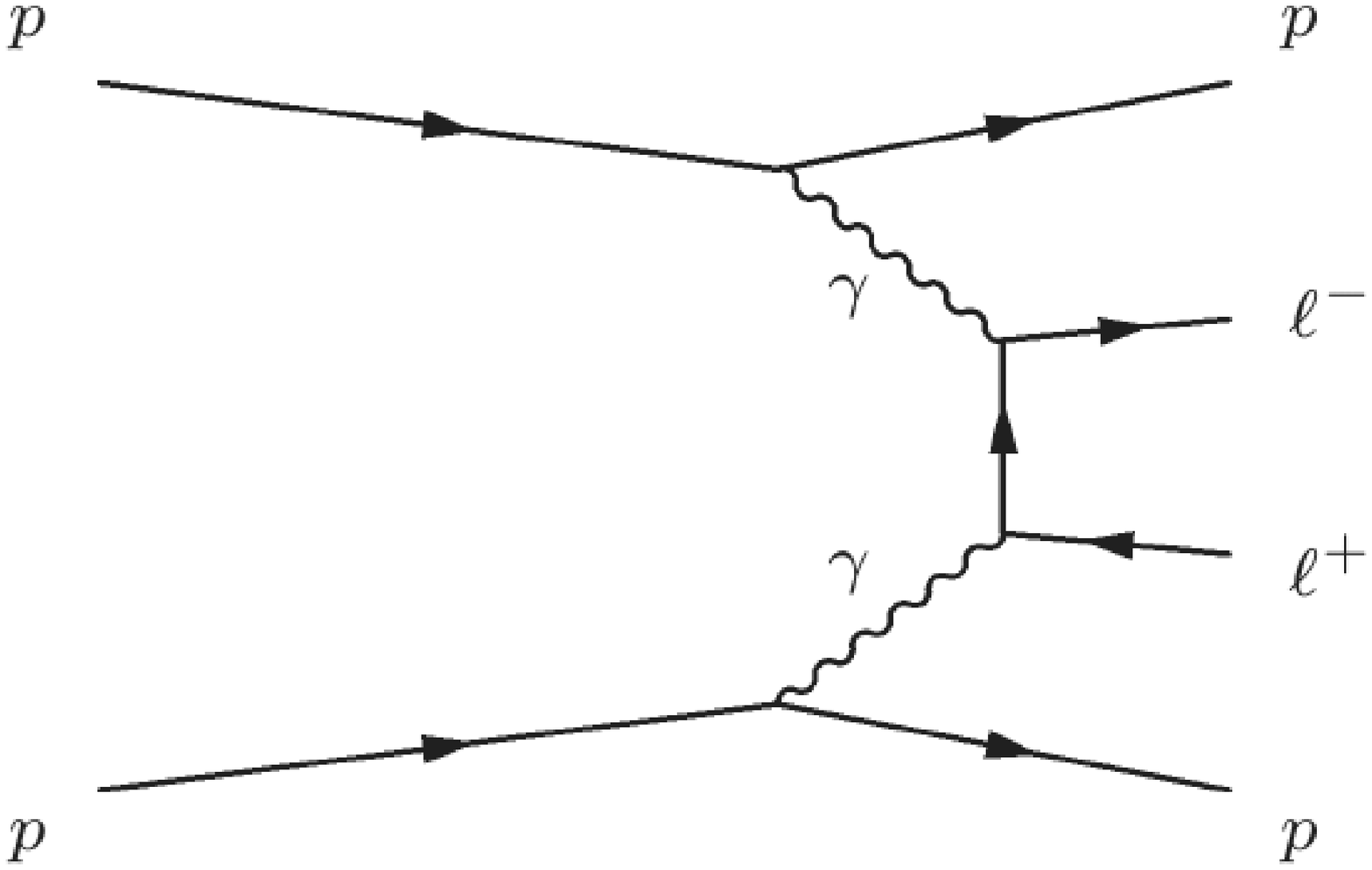}
\includegraphics[width=0.45\columnwidth,height=2.8cm]{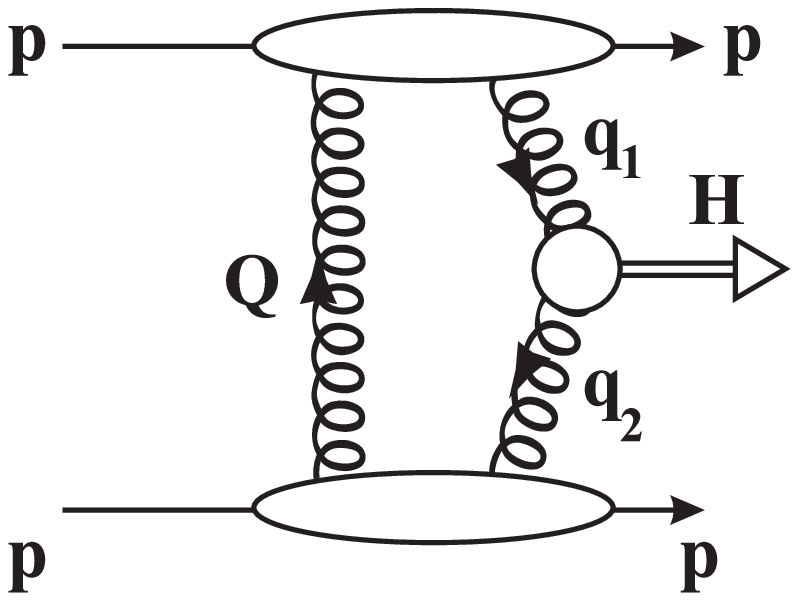}
\caption{Diagrams depicting processes in p-p or Pb-Pb collisions at the LHC 
characterized by forward particle emission: (a) Mueller-Navelet dijets, (b) exclusive quarkonia 
photoproduction, (c) exclusive dileptons, (d) central exclusive Higgs.}
\label{fig:feyn_diags}
\end{figure}

The CERN Large Hadron Collider (LHC) will deliver proton-proton, proton-nucleus and 
nucleus-nucleus collisions at $\sqrtsnn =$~5.5, 8.8 and 14~TeV respectively,
opening up an unprecedented phase-space for particle production spanning up to 
$\Delta\eta\sim$ 20 units of rapidity. Many interesting scattering processes  (mostly 
mediated by colorless exchanges) 
are characterized by particles emitted at very low angles with respect to the beam. Figure~\ref{fig:feyn_diags} 
shows a few representative diagrams mediated respectively by (a) partons, 
(b) a photon and a Pomeron, (c) two photons, and (d) two gluons in a color-singlet state. 
All these processes are characterized by forward particles (jets, protons, ions) 
in the final-state plus an (often exclusive) system produced at more central rapidities.
We present a summary of the physics programme accessible with 
the forward instrumentation capabilities available in the six LHC experiments: 
ALICE, ATLAS, CMS, LHCb, LHCf and TOTEM.

\subsection{Forward kinematics variables}
\label{sec:kinematics}

\begin{figure}[htbp]
\centerline{\includegraphics[width=0.9\columnwidth,height=6.7cm]{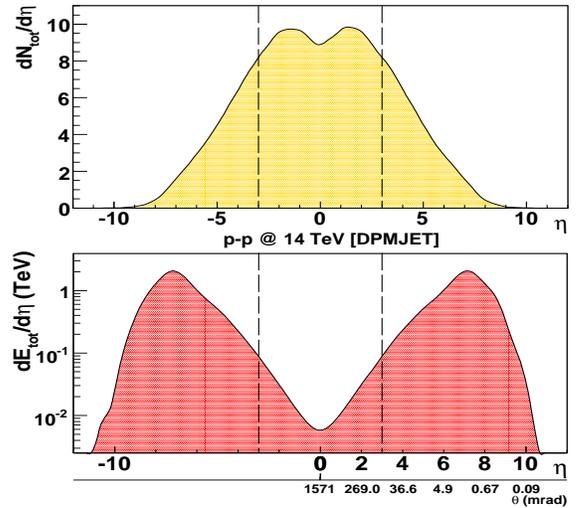}}
\caption{Pseudo-rapidity distributions for the total hadron multiplicity (top) and 
energy (bottom) in p-p at $\sqrts$~=~14 TeV as given by the DPMJET3 model~\cite{dpmjet3}.}
\label{fig:dN_dE_deta}
\end{figure}

The appropriate kinematic variables in (inelastic) hadronic collisions are 
the transverse momentum, $p_T=p \sin\theta$, and the rapidity $y=0.5 \log(E+p_L/E-p_L)$ or $y=\mbox{atanh}(p_L/E)$, 
where $p_L=p\cos\theta$ is the longitudinal momentum, and $\theta$ is the polar angle 
with respect to the beam axis. The rapidity can be thought of as the relativistically-invariant 
measure of the longitudinal velocity. Often the {\it pseudorapidity} $\eta$~=~-ln$\,$tan($\theta/2$) or $\eta=\mbox{atanh}(p_L/p)$, 
which depends solely on $\theta$, is used (note that $y\approx\eta$ for negligible masses i.e. for $E\approx p$, provided that $\theta$ is not very small). 
From Fig.~\ref{fig:dN_dE_deta}, one can see that particle 
production in hadronic collisions is peaked at $\eta$~=~0 (i.e. at 90$^\circ$)~\footnote{Note that $dN/dy$ 
(not shown) has indeed a Gaussian shape with maximum at $y$~=~0. The dip in $dN/d\eta$ is just an 
``artifact'' due to the transformation from rapidity to pseudorapidity ($dy/d\eta$~=~$p/E$ Jacobian).}, 
while most of the 
energy is carried out by particles not far from the beam rapidity, $y_{\rm max}$ = ln($\sqrts/m_p$)~=~9.54 at 14~TeV.
In elastic or diffractive collisions, one deals with particles scattered at very small 
angles and $\eta$ is less useful a variable (ultimately, $\eta\to\infty$ for $\theta\approx$~0). Instead, 
the Feynman $x_F = 2 p_L/\sqrt{s}$ or equivalently $\xi \approx 1 - x_F$, and the 
four-momentum transfer $-t\approx (p\,\theta)^2\approx p_T^2$, are used.\\

\subsection{Forward detectors at the LHC}
\label{sec:dets}

\begin{figure}[htbp]
\centerline{\includegraphics[width=\columnwidth]{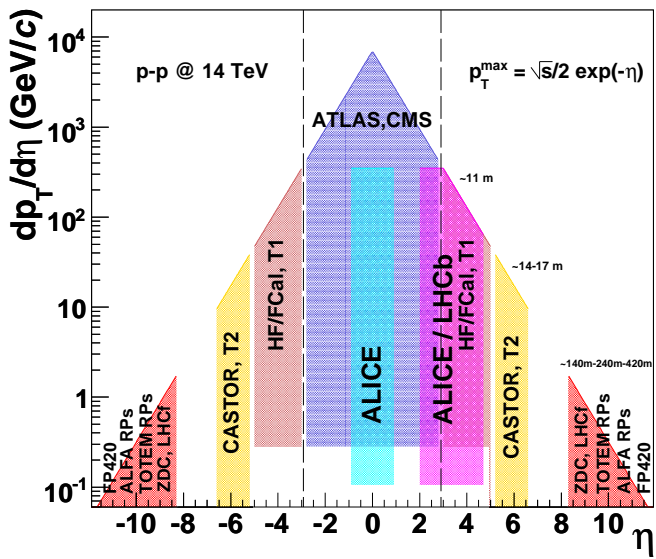}}
\caption{Approximate $p_T$-$\eta$ coverage of current (and proposed) detectors at the LHC 
(adapted from~\cite{risto06}).}
\label{fig:coverage}
\end{figure}

\begin{figure}[htbp]
\centerline{\includegraphics[width=0.95\columnwidth,height=5.cm]{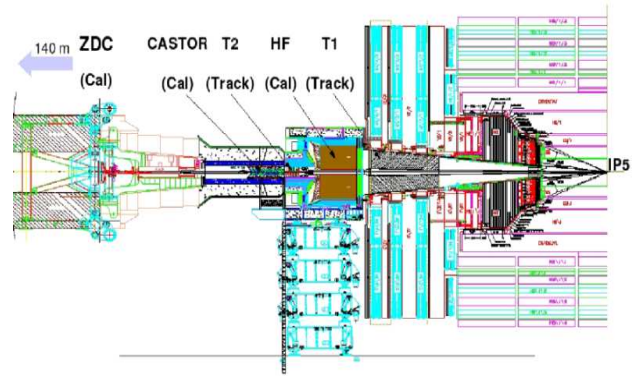}}
\caption{Layout of the detectors in the CMS/TOTEM forward region at the LHC
interaction point 5~\cite{Albrow:2006xt}.}
\label{fig:fwd_cms}
\end{figure}

If one (somewhat arbitrarily) defines ``forward'' rapidities as those beyond $|\eta|\approx$~3, 
all LHC experiments feature forward detection capabilities without parallel
compared to previous colliders (Figs.~\ref{fig:coverage} and~\ref{fig:fwd_cms}):
\begin{itemize}
\item ATLAS~\cite{atlas} and CMS~\cite{cms,Albrow:2006xt} not only cover the largest 
$p_T$-$\eta$ ranges at mid-rapidity for hadrons, electrons, photons and muons, but they 
feature extended instrumentation at distances far away from the interaction point (IP). 
Forward calorimetry is available at $\pm$11 m (the FCal and HF~\cite{hf} hadronic calorimeters), 
at $\pm$14~m (CMS CASTOR sampling calorimeter)~\cite{castor}, and at
$\pm$140~m (the Zero-Degree-Calorimeters, ZDCs)~\cite{zdc_atlas,Grachov:2006ke}.
In addition, ATLAS has (or will have) Roman Pots (RPs) at $\pm$220,240~m~\cite{ATLAS220m,ALFA}, 
and there are advanced plans to install a new proton-tagger system at 420~m (FP420) 
from both the ATLAS and CMS IPs~\cite{fp420}. 
\end{itemize}

\begin{itemize}
\item Both ALICE~\cite{alice} and LHCb~\cite{lhcb} have forward {\it muon} spectrometers 
in regions, 2~$\lesssim~\eta~\lesssim~5$, not covered by ATLAS or CMS. In addition, 
LHCb has good tracking, calorimetry and particle identification 
for the measurement of hadrons, electrons and photons in this $\eta$ range; and ALICE has also ZDCs at $\pm$116 m~\cite{alice_zdc}.
\item The TOTEM experiment~\cite{totem}, sharing IP5 with CMS, features two types 
of trackers (T1 and T2 telescopes) covering $3.1<|\eta|<4.7$ and $5.2<|\eta|<6.5$ 
respectively, plus proton-taggers (Roman Pots) at $\pm$147 and $\pm$220 m.
\item The LHCf experiment~\cite{lhcf} has installed scintillator/silicon calorimeters in the 
same region of the ATLAS ZDCs, $\pm$140 m away from IP1.
\end{itemize}

A rich variety of physics measurements are accessible utilizing such forward instrumentation
in three possible detection modes: 
\begin{enumerate}
\item as detectors to {\it directly measure the 4-momentum} of a given final-state produced 
in the reaction e.g. a jet in CASTOR, a zero-degree photon in LHCf/ZDC, or a leading
proton in FP420;
\item as {\it tagging} devices to signal the presence of a diffractively or elastically 
scattered proton (or a neutron) in Roman Pots (or ZDCs); 
\item as {\it vetoing} devices of final-state particles in the collision,
e.g. requiring no hadronic activity within a given (forward) rapidity range covered by 
one or more detectors.
\end{enumerate}

\subsection{Forward physics at the LHC}
\label{sec:physics}

The following QCD, electro-weak, and Higgs physics topics, developed in more detail in the 
remainder of the document, can be studied with forward instrumentation:\\

\paragraph{QCD physics} (Section~\ref{sec:qcd}).
Many aspects of the physics of the strong interaction can be studied with forward detectors. 
First, Pomeron-induced processes~\cite{Arneodo:2005kd,Goulianos:2007xr} 
-- such as  elastic $p$-$p$ cross section, soft diffractive processes, rapidity-gap survival probability, 
hard diffraction cross sections, etc. -- are accessible with the TOTEM and ATLAS Roman Pots and/or 
by requiring a large enough rapidity gap in one (or both) of the forward hemispheres 
(e.g. HF+CASTOR in CMS). Second, low-$x$ QCD physics ~\cite{cgc,d'Enterria:2006nb} 
-- gluon saturation, non-linear QCD evolution, small-$x$ parton distribution functions (PDFs), 
multi-parton scattering -- can be studied via the measurement of hard QCD cross sections in the 
forward direction (e.g. jets, or direct-$\gamma$ in HF/FCal, CASTOR) or in exclusive photoproduction 
$\gp$~\cite{louvain} and $\gA$~\cite{Baltz:2007kq} processes, tagged with forward 
protons (neutrons) in RPs (ZDCs). Third, hadronic models of ultra-high-energy (UHE) 
cosmic-rays interactions in the upper atmosphere can be effectively tuned by measuring  
the forward energy ($dE/d\eta$) and particle ($dN/d\eta$) flows in $p$-$p$, $p$-$A$, and $A$-$A$ 
collisions~\cite{CRsQM08}.

\paragraph{Electroweak sector} (Section~\ref{sec:ewk}).
All charges accelerated at very high energies generate electromagnetic fields which, 
in the equivalent photon approximation, can be considered as (quasi-real) photon
beams. A significant fraction of the $p$-$p$~\cite{louvain} and $Pb$-$Pb$~\cite{Baltz:2007kq} 
collisions at the LHC will thus involve photon interactions at TeV energies 
giving access to 
a unique programme of $\gamma$-induced studies. 
In particular, photon-photon interactions, tagged with forward protons (neutrons) 
in the RPs (ZDCs), allow one to measure the beam luminosity 
via the pure QED process $\gaga \rightarrow \lele$; or to study (anomalous) gauge 
boson couplings via $\gp, \gA \rightarrow p \,n\, W$, or $\gaga \rightarrow ZZ,\,WW$.

\paragraph{Higgs physics} (Section~\ref{sec:higgs}).
Two mechanisms of Higgs production at the LHC are accompanied by forward particle 
emission. First, the vector-boson-fusion (VBF) process,  $pp \rightarrow qq \rightarrow q H q$, 
where the two valence quarks radiate $W$ or $Z$ bosons which merge to form the Higgs, and then 
fragment into two forward-backward jets tagged in the forward calorimeter systems.
Second, the central exclusive channel, $pp \rightarrow p H  p$ (diagram (d) of Fig.~\ref{fig:feyn_diags}),
where the Higgs boson is produced at central rapidities from the fusion of a two-gluon 
(color-singlet) system~\cite{DeRoeck:2002hk} and the interacting protons, scattered intact at very small angles,
are measured e.g. in the planned FP420 proton spectrometer~\cite{fp420}.

\section{QCD physics}
\label{sec:qcd}

\subsection{Elastic scattering}

\begin{figure}[htbp]
\hspace{-1.1cm}\centering\includegraphics[width=0.9\columnwidth,height=4.6cm]{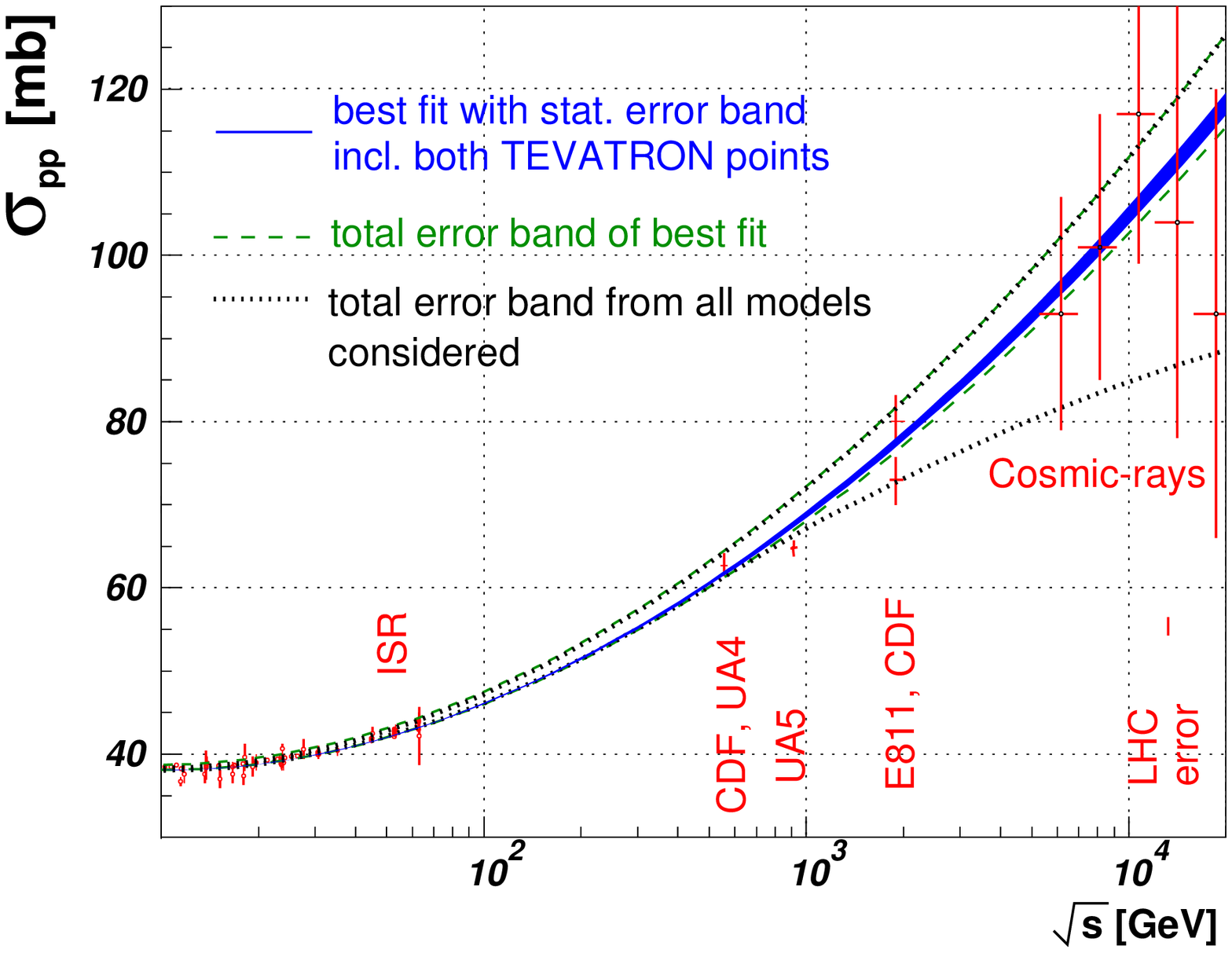}
\centering\includegraphics[width=0.9\columnwidth,height=5.5cm]{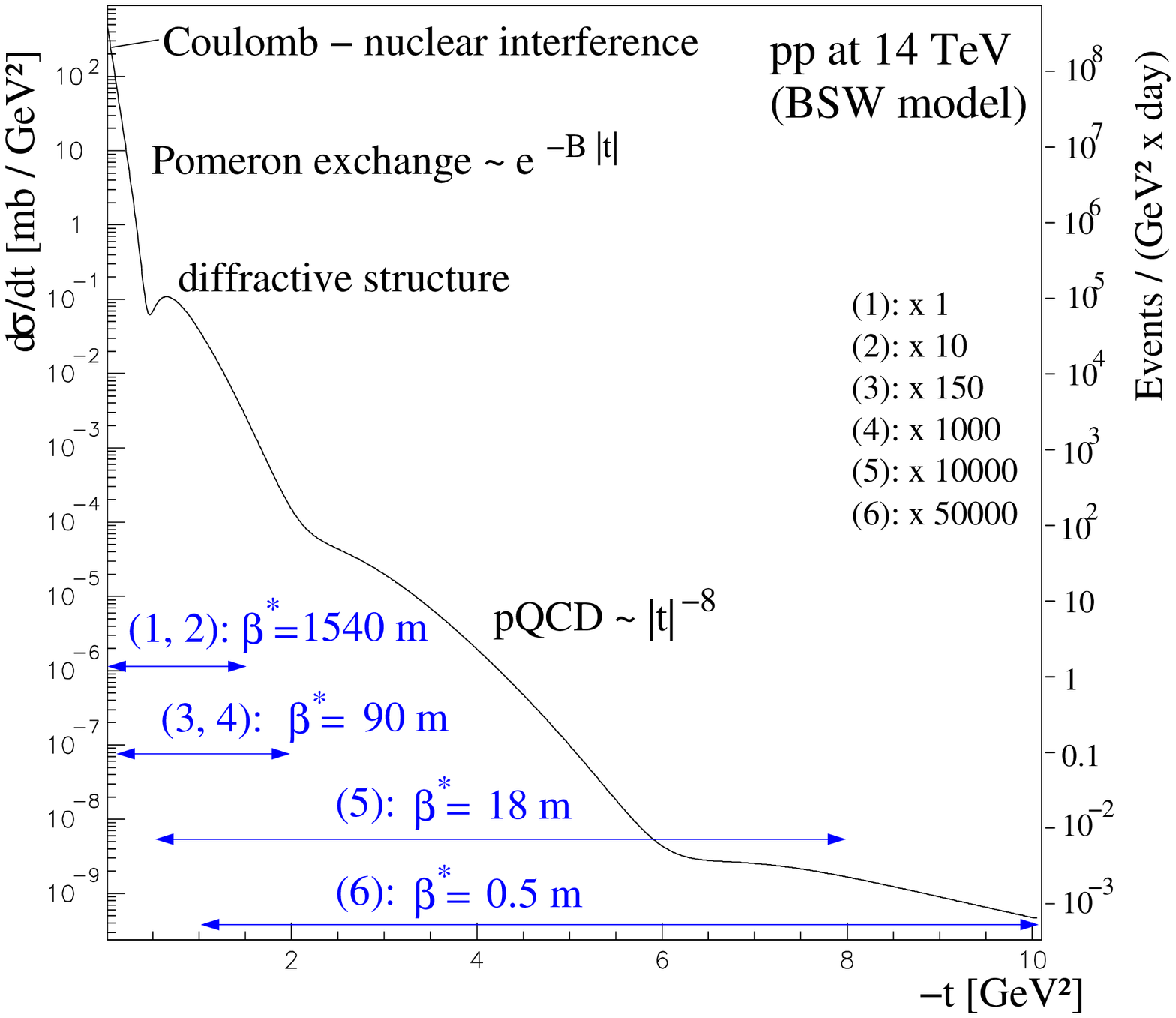}
\caption{Top: COMPETE predictions~\protect\cite{compete} for $\sigma_{tot}$ with statistical 
(blue solid) and total (dashed) errors (including the Tevatron ambiguity) compared to existing
data. Bottom: Prediction for elastic p-p scattering at the LHC with 
various beam optics settings~\protect\cite{Anelli:2006cz}.}
\label{fig:totem}
\end{figure}

The measurement at the LHC of the total $p$-$p$ cross section and of the $\rho$-parameter 
(the ratio of real to imaginary part of the forward elastic scattering amplitude) 
provides a valuable test of fundamental quantum mechanics relations such as the Froissart bound 
$\sigma_{tot}<${\small{\it Const}} $\ln^2s$, the optical theorem $\sigma_{tot}\sim $Im$f_{el}(t=0)$, 
and dispersion relations Re$f_{el}(t=0)\sim$Im$f_{el}(t=0)$~\cite{Bourrely:2005qh}. 
The current extrapolations of the total $p$-$p$ cross section at the LHC ($\sigma_{tot}$ = 90 -- 140~mb),
of which the elastic contribution accounts for about one fourth, suffer from large uncertainties due to a 
2.6$\sigma$ disagreement between the E710 and CDF measurements at Tevatron 
(Fig.~\ref{fig:totem}, top).\\

The main goal of TOTEM experiment is to obtain a precise measurement of the total and 
elastic $p$-$p$ cross sections over a large range of 4-momentum transfers from 
$-t\approx 2 \cdot 10^{-3}\,$GeV$^{2}$ to 8\,GeV$^{2}$, using different $\beta^{*}$ optics
settings (Fig.~\ref{fig:totem}, bottom). The total $p$-$p$ cross section and the LHC luminosity 
will be measured making use of the optical theorem via
\begin{equation}
\sigma_{tot} = \frac{16 \pi}{1 + \rho^{2}} \cdot
\frac{dN_{el}/dt |_{t=0}}{N_{el} + N_{inel}}\:,\;\;
\qquad
\mathcal{L} = \frac{1 + \rho^{2}}{16 \pi} \cdot 
\frac{(N_{el} + N_{inel})^{2}}{dN_{el}/dt |_{t=0}}. \nonumber
\end{equation}
Assuming an uncertainty in $\rho = 0.12 \pm 0.02$, $\sigma_{tot}$ and 
$\mathcal{L}$ can be measured within about 1\,\%.
Also ATLAS plans to measure the elastic cross section with its Roman Pots~\cite{ALFA} 
based on a fit of the data in the Coulomb region to 
\begin{eqnarray}
\frac{dN}{dt} (t \rightarrow 0) = L \pi \left( \frac{-2 \alpha}{|t|} +
\frac{\sigma_{tot}}{4 \pi} (i+\rho)e^{-b|t|/2} \right)^2.
\end{eqnarray}
Such a measurement requires to go down to $-t \sim 6.5\cdot 10^{-4}$~GeV$^2$
(i.e. $\theta \sim 3.5~\mu$rad) to reach the kinematical domain where the strong
amplitude equals the electromagnetic one. 

\subsection{Hard and soft diffractive processes}

\begin{figure}[htbp]
\centerline{\includegraphics[width=\columnwidth,height=2.9cm]{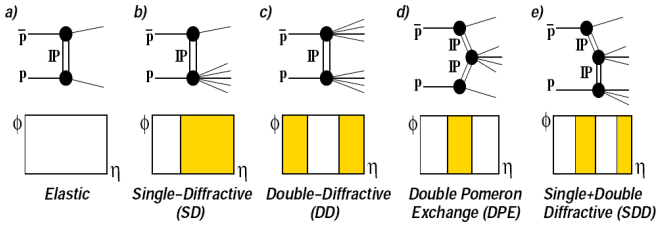}}
\caption{Event topologies in $\eta$ vs azimuth $\phi$ for elastic and diffractive
p-p interactions. Shaded (empty) areas represent particle-production (rapidity-gaps) 
regions~\cite{Goulianos:2007xr}.}
\label{fig:diffract_diags}
\end{figure}

Diffractive physics covers the class of inelastic interactions that contain large rapidity gaps 
(LRGs, $\Delta\eta\gtrsim$ 4) devoid of hadronic activity and where one or both protons
emerge intact in the final state~\cite{Arneodo:2005kd}. Such event topologies, with reduced QCD radiation, imply 
colorless exchange mediated by two or more gluons in a color-singlet state (a {\em Pomeron}, $\pom$). 
Depending on the number and relative separation of the LRGs, one further differentiates 
between single, double, or double-Pomeron-exchange (DPE) processes (Fig.~\ref{fig:diffract_diags}).
The centrally produced system has a mass equal to $M^2 \approx s \xi_1 $ 
($M^2 \approx s \xi_1 \xi_2$) for single diffractive (DPE) events, and the size of the 
rapidity gap is of the order of $\Delta \eta \sim \log 1/ \xi_{1,2}$.\\


The interest of diffractive processes is manifold. On the one hand, soft diffraction processes, 
dominated by non-perturbative (Regge) dynamics, constitute a significant fraction ($\sim$20\%) 
of the total inelastic $p$-$p$ cross section. Their characterization -- in particular the 
so-called ``rapidity-gap survival probability'' -- is important in order to have under control 
the backgrounds of many processes at the LHC~\cite{Ryskin:2007qx}. 
On the other, hard diffraction processes which involve the production of a high-mass 
or large-$p_T$ state ($Q\overline{Q}$, jets, $W$, $Z$ ...) are in principle perturbatively 
calculable and provide information on diffractive (or generalized) Parton Distribution 
Functions, dPDFs (GPDs), which describe not only the density of partons in the proton 
but also their correlations~\cite{Diehl:2003ny}.

\subsection{Parton structure and evolution at low-$x$}

\begin{figure}[htbp]
\centerline{\includegraphics[width=0.8\columnwidth,height=7.cm]{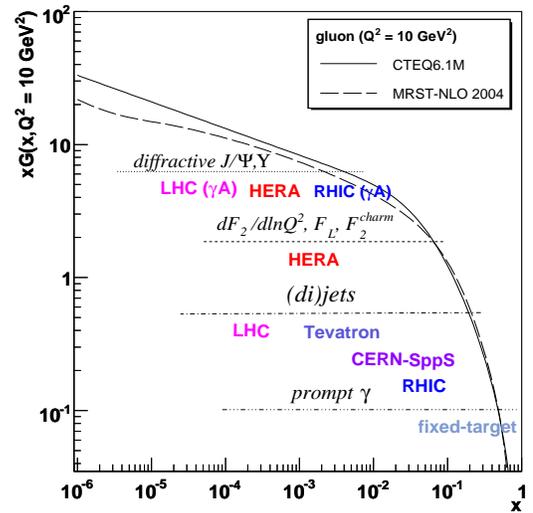}}
\caption{Examples of measurements providing information on the gluon PDF as a function of $x$~\cite{d'Enterria:2006nb}.}
\label{fig:HERA_xG}
\end{figure}

Figure~\ref{fig:HERA_xG} summarizes the methods at hand to determine 
the gluon density $xG(x,Q^2)$ in the proton as a function of fractional momenta 
$x=p_{\mbox{\tiny{\it parton}}}/p_{\mbox{\tiny{\it proton}}}$. 
The main source of information so far on $xG(x,Q^2)$ is (indirectly) obtained 
from the $\ln Q^2$ slope (``scaling violations'') of the $F_2$ structure function
in $e$-$p$ deep-inelastic-scattering (DIS). Additional constraints can be obtained from 
$F_2^{charm}$~\cite{hera_lhc_heavyQ} and diffractive photoproduction of 
heavy vector mesons ($\jpsi,\ups$)~\cite{teubner07} and, in particular 
(since $xG\propto F_L$), from the longitudinal structure function $F_L$. 

\begin{figure}[htbp]
\vspace{.3cm}
\centering
\includegraphics[width=0.8\columnwidth,height=4.8cm]{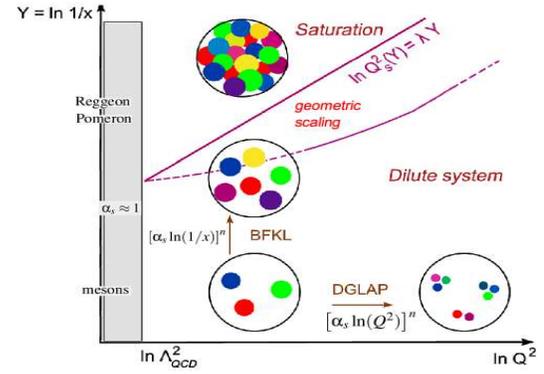}
\caption{log(1/$x$)-Q$^2$ plane with the different QCD evolution regimes (DGLAP, BFKL, saturation).}
\label{fig:cgc}
\vspace*{-.5cm}
\end{figure}

For decreasing parton momentum fraction $x$, the gluon density is observed to grow rapidly.
As long as the densities are not too high, this growth is described by the 
Dokshitzer-Gribov-Lipatov-Altarelli-Parisi (DGLAP)~\cite{dglap} or by the 
Balitski-Fadin-Kuraev-Lipatov (BFKL)~\cite{bfkl} evolution equations which govern, 
respectively, parton radiation in $Q^2$ and $x$ (Fig.~\ref{fig:cgc}). 
Eventually, at high enough center-of-mass energies (i.e. at very small $x$) the gluon 
density will be so large that non-linear (gluon-gluon fusion) effects will become important
~\cite{sat}. A regime of saturated parton densities is thus expected for small enough $x$ 
values at virtualities below an energy-dependent ``saturation momentum'' scale, $Q_s$, 
intrinsic to the size of the hadron. Saturation effects are amplified in nuclear targets 
because of their increased transverse parton density compared to the proton (for nuclei, 
$Q_s^2\propto A^{1/3}$ where $A\approx$~200 is the mass number in $Pb$ or $Au$)~\cite{cgc}.\\ 

In hadron-hadron collisions, information on $xG$ can be obtained in processes with prompt photons, 
jets, and heavy-quarks in the final state. In a $2\rightarrow 2$ parton scattering 
the {\it minimum} momentum fraction probed when a particle of momentum $p_T$ is
produced at pseudo-rapidity $\eta$ is $x_{min}=2p_T/\sqrt{s}\,exp(-\eta)$, i.e. $x_{min}$ 
decreases by a factor of $\sim$10 every 2 units of rapidity. Thus, forward instrumentation 
provides an important lever arm for the measurement of the low-$x$ structure 
and evolution of the parton densities.
Three representative low-$x$ QCD measurements at the LHC~\cite{d'Enterria:2006nb} are discussed next.

\subsubsection*{$\bullet$ Case study I: Forward (di)jets}

\begin{figure}[htbp]
\hspace{-0.2cm}\centering\includegraphics[width=0.83\columnwidth,height=6.cm]{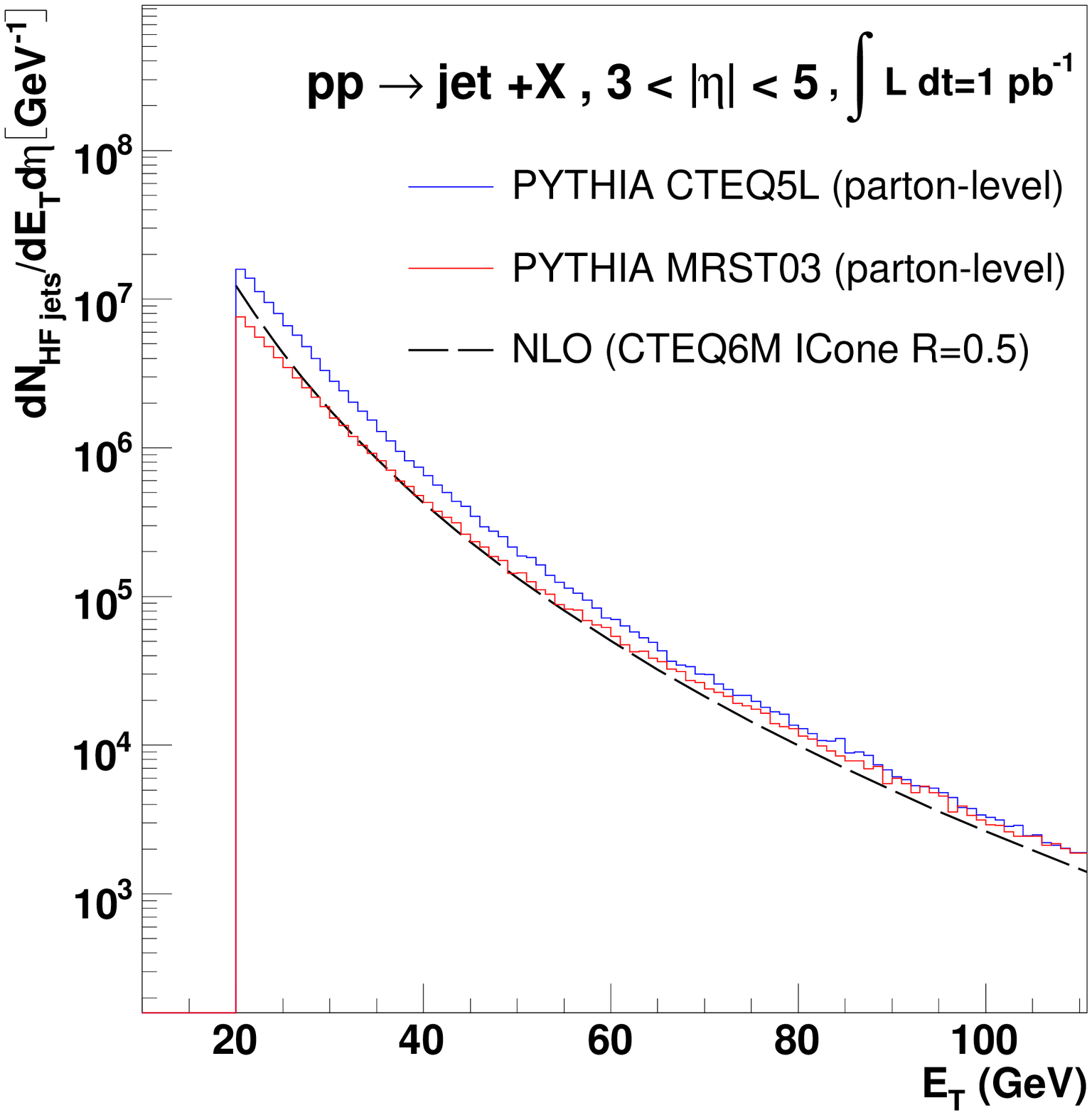}
\centering\includegraphics[width=0.82\columnwidth,height=5.8cm]{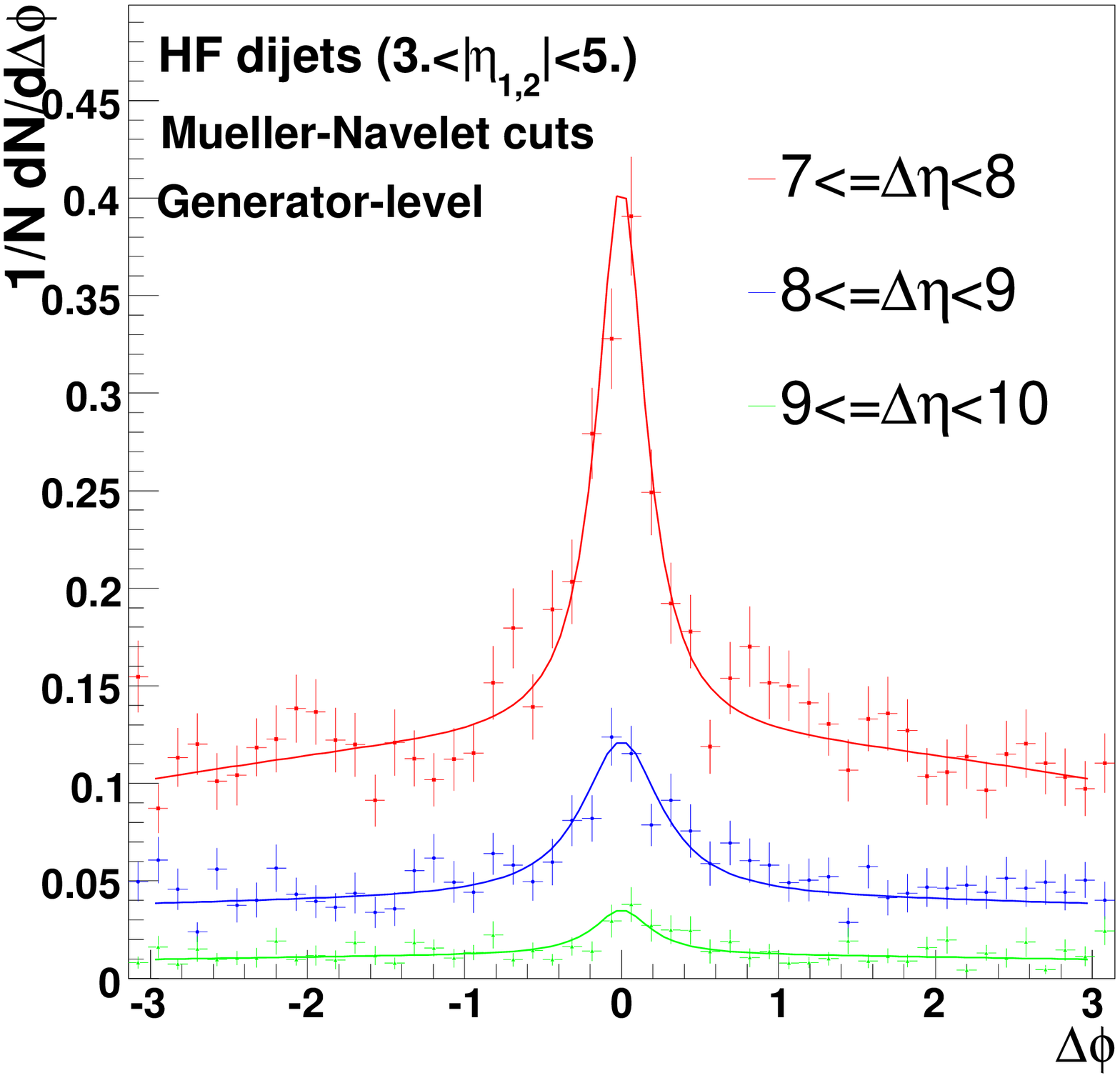}
\caption{Forward jet results in  (1~pb$^{-1}$) $p$-$p$ collisions at $\sqrts$~=~14~TeV~\protect\cite{scerci_dde}. 
Top: Single inclusive jet yields in the CMS forward calorimeters expected with various PDF sets. 
Bottom: $\Delta\phi=(\phi_{1}-\phi_{2})-\pi$ distributions for dijet events passing the 
Mueller-Navelet cuts with separations $\Delta\eta$~=~7.5, 8.5 and 9.5.}
\label{fig:fwd_jets}
\end{figure}

From the formula for $x_{min}$ above, it follows that the measurement of relatively 
soft jets with $E_T\approx$~20--100~GeV in the forward calorimeters (3$<|\eta|<$6.6) 
allows one to probe the PDFs at $x$ values as low as $10^{-5}$, in $p$-$p$ at 14~TeV. 
In this kinematic range, the current PDF uncertainties result in variations of the jet cross sections as large
as 60\% (Fig.~\ref{fig:fwd_jets}, top). The interest in forward jet measurements goes beyond the {\it single} 
inclusive cross sections: the production of {\it dijets} with similar $E_T$ but separated by large
rapidities, the so-called ``Mueller-Navelet jets'' (diagram (a) of Fig.~\ref{fig:feyn_diags})~\cite{mueller_navelet}, 
is a particularly sensitive measure of BFKL~\cite{DelDuca93,sabiovera} 
as well as non-linear~\cite{marquet,iancu08} QCD evolutions. The large rapidity interval 
between the jets (e.g. up to $\Delta\eta \approx$~12 in the extremes of CMS forward calorimeters) 
enhances large logarithms of the type $\Delta\eta \sim log(s/E_{T,1}E_{T,2})$ which can be appropriately 
resummed using the BFKL equation. One of the phenomenological implications of BFKL dynamics
is an enhanced radiation between the two jets which results in a larger azimuthal 
decorrelation for increasing $\Delta\eta$ separations, compared to collinear pQCD approaches. 
Preliminary CMS analyses~\cite{scerci_dde} indicate that such studies are well feasible 
by measuring jets in each one of the HF forward calorimeters (Fig.~\ref{fig:fwd_jets}, bottom).

\subsubsection*{$\bullet$ Case study II: Forward heavy-quarks}
\label{sec:heavyQ}

\begin{figure}[htbp]
\hspace*{-0.3cm}
\includegraphics[width=0.87\columnwidth,height=5.7cm]{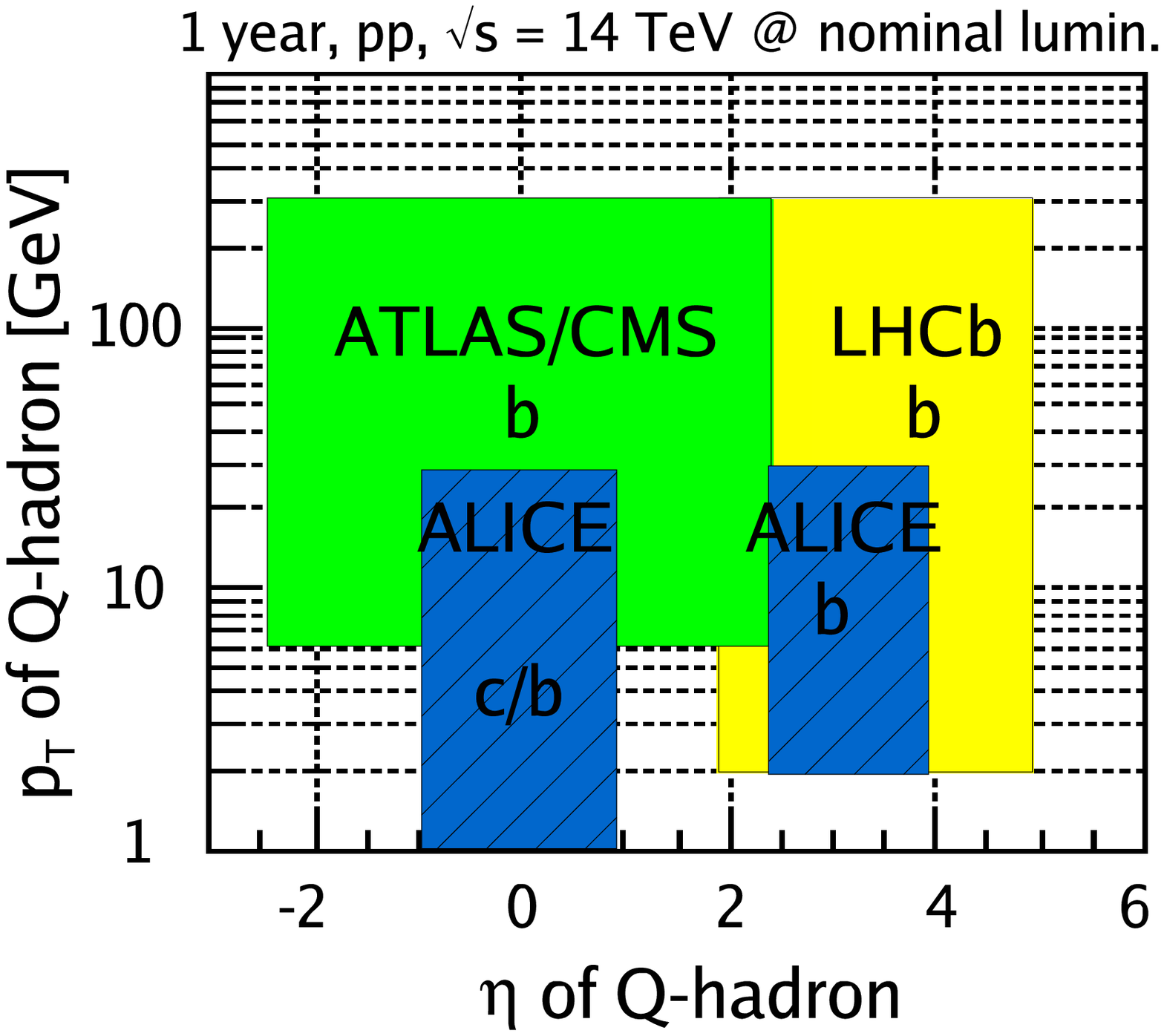}\vspace{0.2cm}
\hspace{0.4cm}\includegraphics[width=0.85\columnwidth,height=5.1cm]{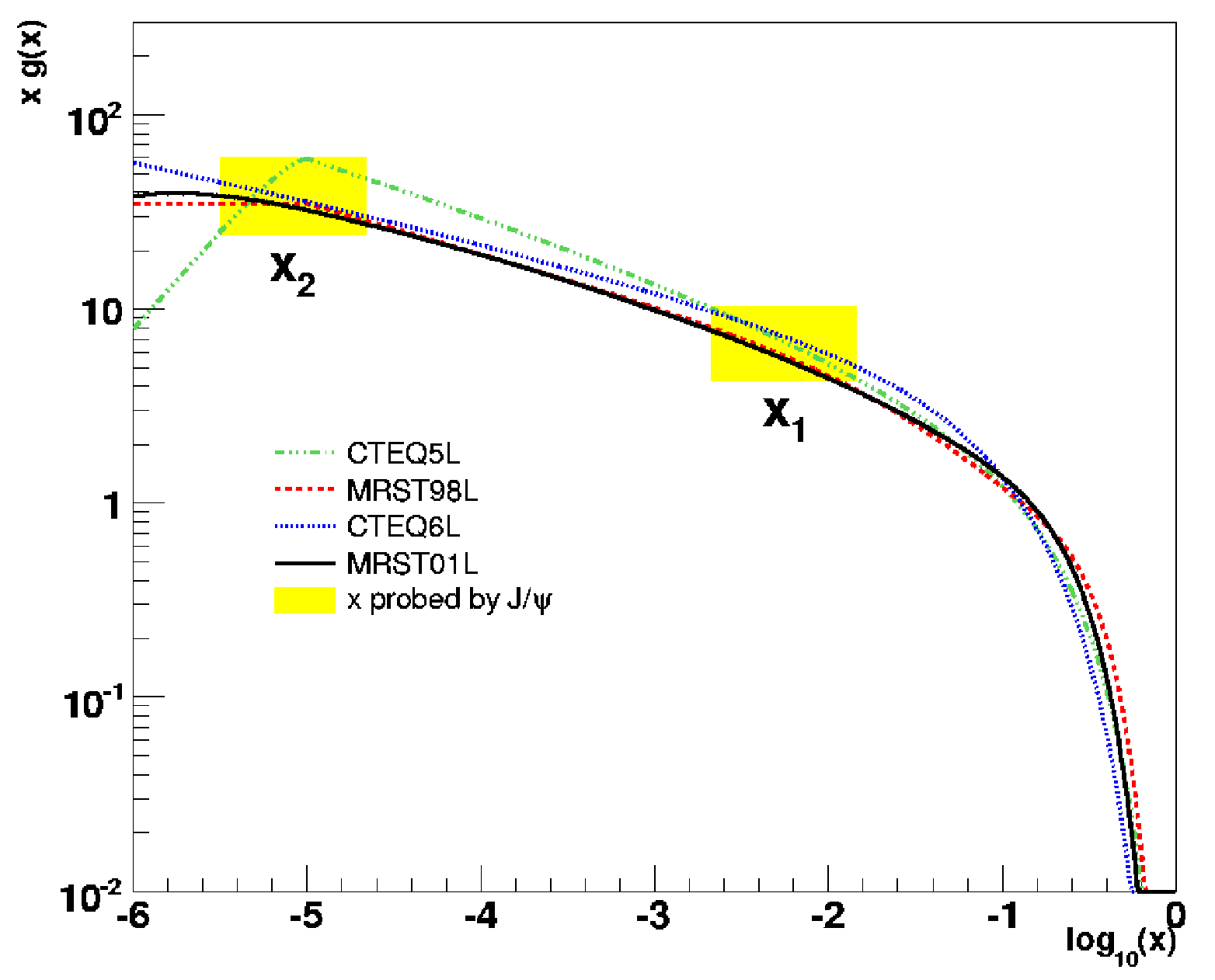}
\caption{Top: Acceptances in $(\eta,p_T)$ for open charm and bottom at the LHC~\cite{hera_lhc_heavyQ}. 
Bottom: Sensitivity of the forward $\jpsi$ measurement in ALICE to various $x$ ranges of the gluon PDF~\cite{stocco}.}
\label{fig:heavyQ_QQbar_LHC}
\end{figure}

The possibility of ALICE and LHCb (Fig.~\ref{fig:heavyQ_QQbar_LHC}, top) to reconstruct 
heavy $D$ and $B$ mesons as well as quarkonia in a large forward rapidity range can also add
stringent constraints on the gluon structure and evolution at low-$x$. Studies of small-$x$ 
effects on heavy flavor production based on collinear and $k_T$ factorization, including 
non-linear terms in the parton evolution, lead to varying predictions for the measured $c$ 
and $b$ cross sections at the LHC~\cite{hera_lhc_heavyQ}. The hadroproduction of $\jpsi$ 
proceeds mainly via gluon-gluon fusion and, having a mass around the saturation scale 
$Q_s\approx$~3~GeV at the LHC, is also a sensitive probe of possible gluon saturation phenomena. 
Figure~\ref{fig:heavyQ_QQbar_LHC} (bottom) shows the gluon $x$ range probed in $p$-$p$ collisions 
producing a $\jpsi$ inside the ALICE muon arm acceptance ($2.5\lesssim \eta \lesssim 4$). 
The observed differences in the underlying PDF fits translate into variations as large as 
a factor of $\sim$2 in the finally measured cross sections~\cite{stocco}.

\subsubsection*{$\bullet$ Case study III: Q$\overline{Q}$ exclusive photoproduction}
\label{sec:upc_qqbar}

\begin{figure}[htbp]
\centering
\includegraphics[width=0.9\columnwidth,height=6.3cm]{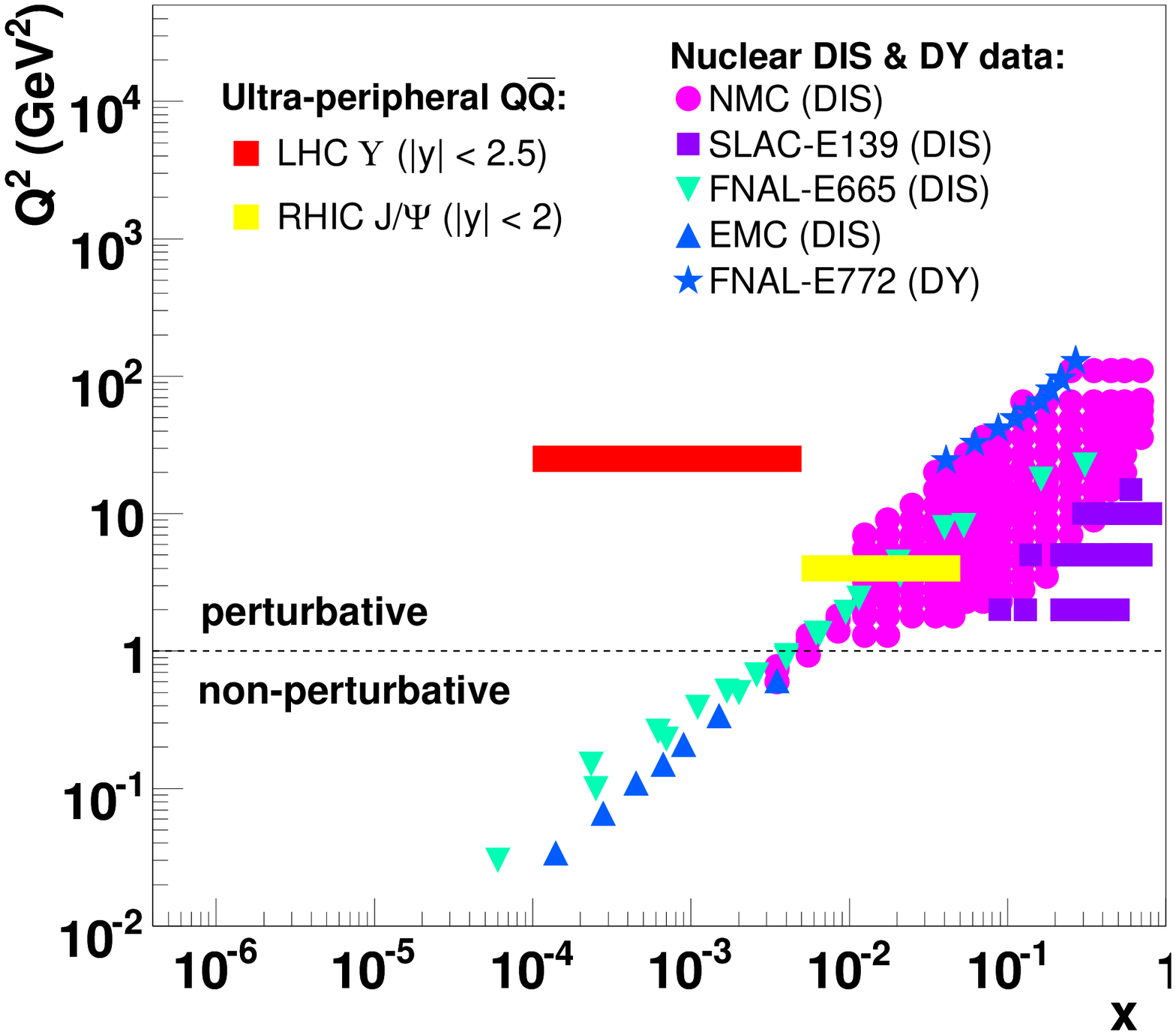}
\includegraphics[width=0.87\columnwidth,height=5cm]{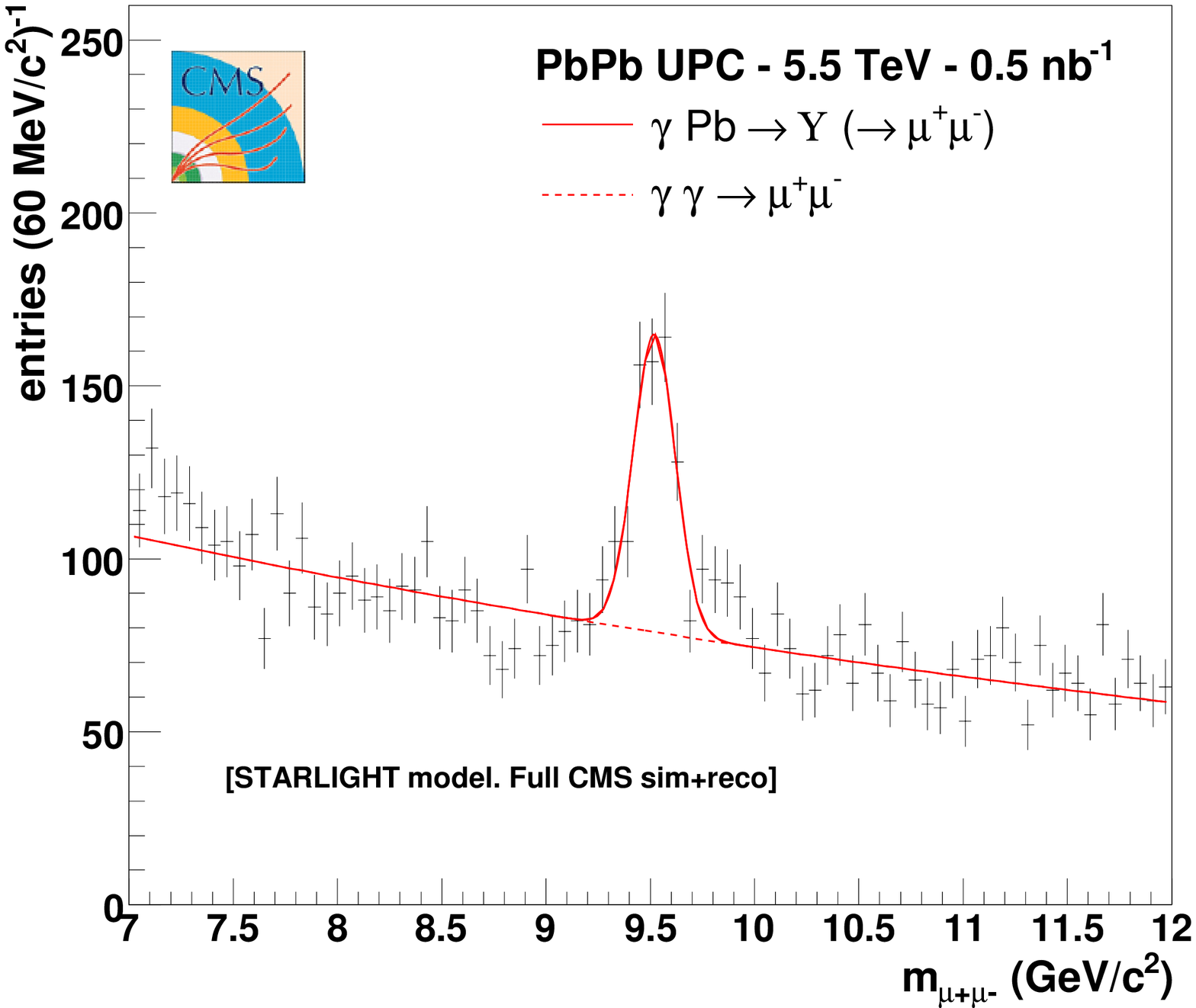}
\caption{Top: Kinematic ($x,Q^2$) plane probed in e-,$\gamma$-A collisions: 
DIS data compared to ultraperipheral $\qqbar$ photoproduction ranges~\protect\cite{d'Enterria:2007pk}.
Bottom: Expected dimuon invariant mass from $\gamma\, Pb \rightarrow \Upsilon\, Pb^\star$ 
on top of the $\gaga\rightarrow \mumu$ continuum in ultraperipheral $Pb$-$Pb$ collisions at $\sqrtsnn =$~5.5~TeV~\protect\cite{cms_hi_ptdr}.}
\label{fig:upc_ups_cms}
\end{figure}

High-energy photons from electromagnetic (ultraperipheral) proton-proton or ion-ion interactions
can be used to constrain the low-$x$ behavior of the nuclear gluon density via exclusive 
photoproduction of quarkonia, dijets and other hard processes~\cite{Baltz:2007kq}.
This is particularly interesting with heavy-ions where, thanks to the large nuclear charge 
($Z$~=~80 for Pb), the available photon fluxes ($dN_\gamma/d\omega\propto Z^2$)
allow one to precisely probe the barely known nuclear gluon distribution (Fig.~\ref{fig:upc_ups_cms}, top).
Lead beams at 2.75~TeV have Lorentz factors $\gamma$ = 2930 leading to maximum (equivalent) 
photon energies $\omega_{\ensuremath{\it max}}\approx \gamma/R\sim$ 100~GeV,
and corresponding maximum c.m. energies: 
$W_{\gaga}^{\ensuremath{\it max}}\approx$~160~GeV and $W^{\ensuremath{\it max}}_{\gA}\approx$~1~TeV,
i.e. 3--4 times higher than equivalent photoproduction studies at HERA.
The $x$ values probed in $\gA \rightarrow Q \overline{Q}\;A$ processes (diagram (b) of Fig.~\ref{fig:feyn_diags})
can be as low as $x\sim 10^{-5}$~\cite{d'Enterria:2007pk}. ALICE, ATLAS and CMS
can measure the $\jpsi,\ups\rightarrow e^+e^-,\mu^+\mu^-$ produced in electromagnetic 
$Pb$-$Pb$ collisions tagged with neutrons detected in the ZDCs. 
Full simulation analyses~\cite{cms_hi_ptdr} indicate that CMS can measure 
a total yield of $\sim$\,500 $\ups$'s within $|\eta|<$ 2.5 for the  nominal 0.5~nb$^{-1}$ $Pb$-$Pb$ 
integrated luminosity (Fig.~\ref{fig:upc_ups_cms}, bottom). With such statistics, studies of the $p_T$ 
and $\eta$ distributions of the $\ups$ can be carried out which will help constrain the low-$x$ gluon 
density in the $Pb$ nucleus. Similar exclusive $\ups$ studies have been conducted in $p$-$p$ collisions~\cite{hollar}.

\subsection{Validation of QCD Monte Carlos for UHE cosmic-rays}

\begin{figure}[htbp]
\centering\includegraphics[width=0.9\columnwidth,height=5.cm]{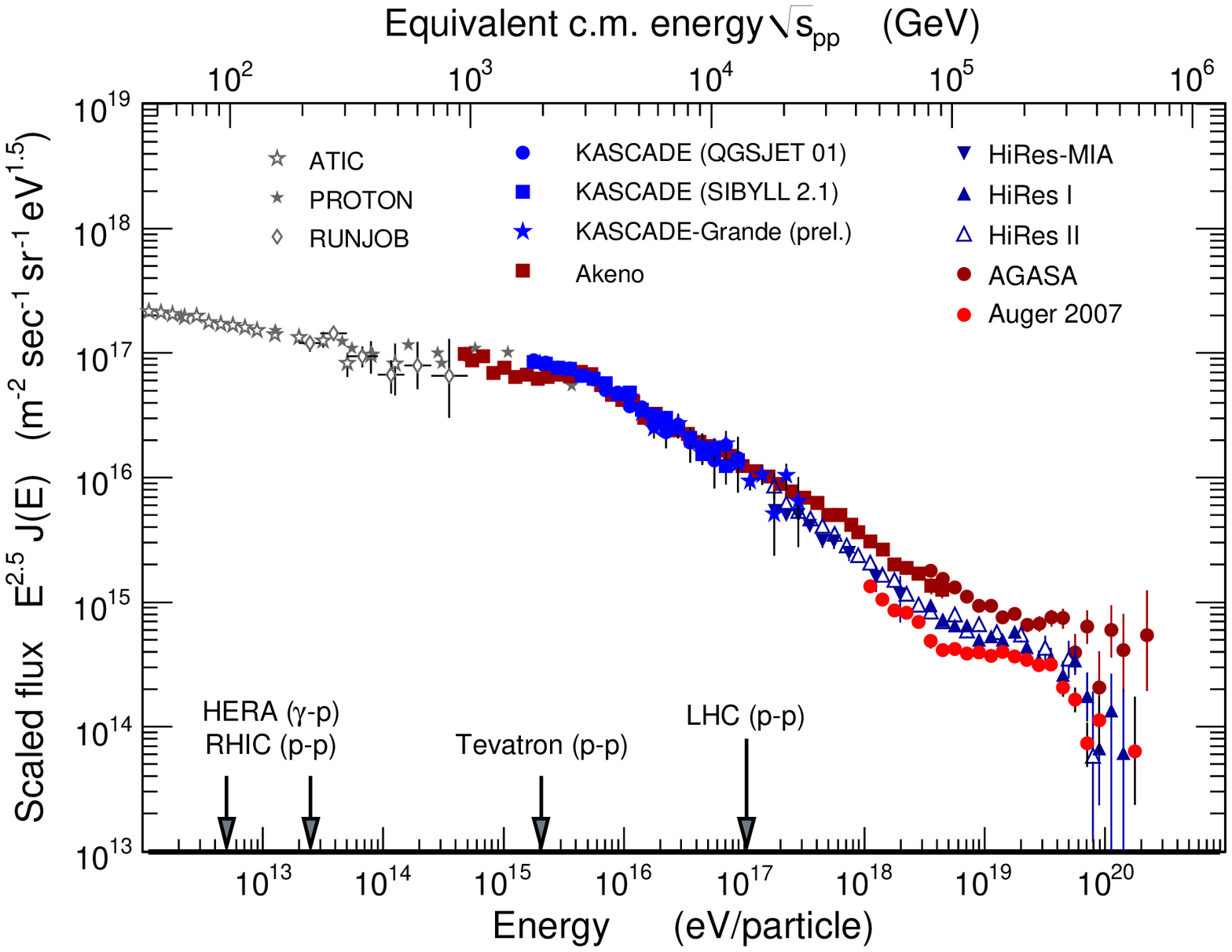}
\centering\hspace{0.9cm}\includegraphics[width=0.92\columnwidth,height=4.9cm]{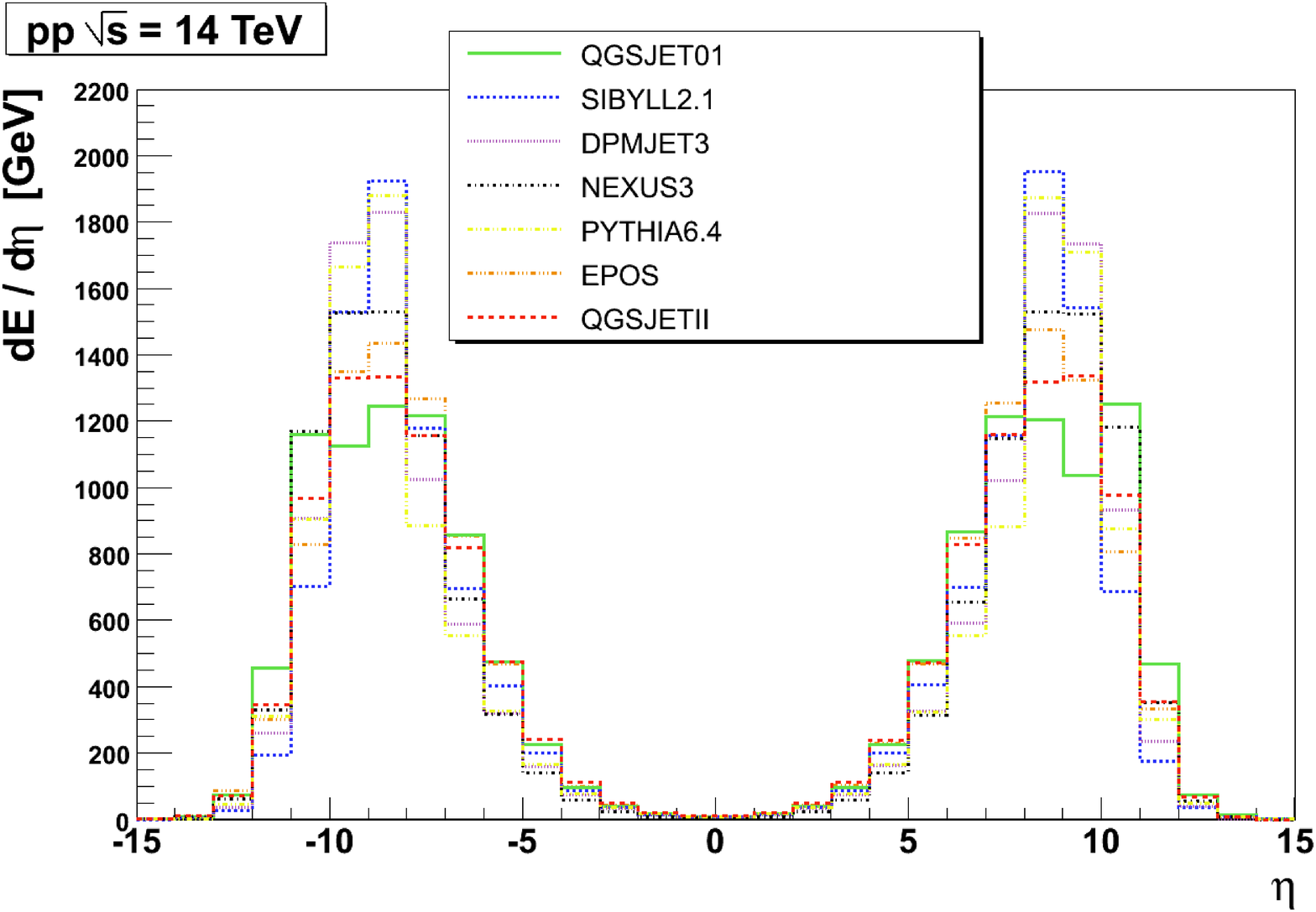}
\caption{Top: Measured CR energy spectrum. 
Bottom: Pseudorapidity energy density for $p$-$p$ at $\sqrts$~=~14~TeV predicted 
by various commonly used MC models in UHE cosmic rays physics~\cite{CRsQM08}.}
\label{fig:cosmics}
\end{figure}

The LHC will not only address fundamental open questions in particle physics but will also provide
valuable insights in closely related domains, such as on the origin and nature of cosmic rays (CRs) 
with energies between $10^{15}$~eV and the so-called ``GZK-cutoff'' at $10^{20}$\,eV,
recently measured by the HiRes~\cite{hires} and Auger~\cite{auger} experiments (Fig.~\ref{fig:cosmics}, top).
CR candidates are protons and nuclei as massive as iron, which generate ``extended air-showers''  
in proton-nucleus ($p$-Air) and nucleus-nucleus ($\alpha$-,Fe- Air) collisions when entering the atmosphere. 
Determination of the primary energy and mass relies on hadronic Monte Carlo codes which 
describe the interactions of the primary (dominated by forward and soft QCD interactions)
in the upper atmosphere~\cite{CRsQM08}. Existing MC models predict energy and multiplicity flows 
differing by factors as large as three, with significant inconsistencies in the forward region. 
The measurement of forward particle production in $p$-$p$, $p$-$A$ and $A$-$A$ 
collisions~\footnote{Note that CRs interactions in the atmosphere are mostly proton-nucleus 
($p$-Air) and nucleus-nucleus ($\alpha$-,Fe- Air) collisions.} at LHC energies (equivalent to 
$E_{lab} \approx 10^{17}$ eV) will provide strong constraints on these models and allow 
for more reliable extrapolations of the CR energy and composition around the GZK cut-off. 
Figure~\ref{fig:cosmics} (bottom) compares the predictions of QGSJET~\cite{qgsjet}, 
DPMJET~\cite{dpmjet3}, NEXUS~\cite{nexus}, EPOS~\cite{epos}, and PYTHIA~\cite{pythia}
for the energy flow ($dE/d\eta$) in $p$-$p$ collisions at $\sqrts$~=~14~TeV. In the range covered 
by detectors like CASTOR or TOTEM (around $|\eta|\approx$~6) and ZDC or LHCf 
(beyond $|\eta|\approx$~8, for neutrals), the model predictions differ by up to $\sim$60\%.

\section{Electroweak physics}
\label{sec:ewk}

\subsection{Beam luminosity via exclusive dileptons}

\begin{figure}[htbp]
\centering
\includegraphics[width=0.87\columnwidth,height=5.5cm]{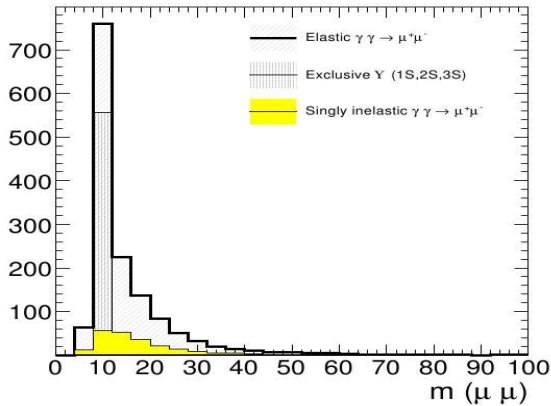}
\caption{Dimuon mass distribution in photon-induced events (for 100$^{-1}$~pb in 
$p$-$p$ at $\sqrts$~=~14~TeV) passing the exclusivity cuts discussed in~\cite{hollar}.}
\label{fig:gg_gp}
\end{figure}

Two-photon dilepton production, $pp \rightarrow p\;\lele p$ (diagram (c) of Fig.~\ref{fig:feyn_diags}) 
is a useful luminosity calibration process, thanks to its precisely known 
(nearly pure) QED cross section~\cite{Bocian:2004ev}. Experimentally, such a process can be 
tagged with rapidity-gaps (``exclusivity'') conditions~\cite{hollar} or via forward protons~\cite{louvain},
and has a clear signature in the exclusive back-to-back dileptons ($|\Delta\phi(\lele)|>2.9$) 
measured within $|\eta|<$~3. The {\sc lpair}~\cite{lpair} cross section for events 
where both muons have $p_T >$ 3~GeV/c and can, therefore, reach the CMS muon chambers is about 50 pb. 
About 710 exclusive dimuons per 100 pb$^{-1}$ are expected in CMS after selection cuts~\cite{hollar}.
The resulting invariant mass distribution 
is shown in Fig.~\ref{fig:gg_gp}, together with the background muons from the $\ups$ decay and inelastic (or dissociative) dimuon events.
This measurement is also easily accessible in electromagnetic $Pb$-$Pb$ collisions 
(see Fig.~\ref{fig:upc_ups_cms} bottom) where the dilepton continuum 
(thanks to the $Z^4$ photon flux enhancement factor) is much larger than in $p$-$p$.\\

\subsection{New physics via anomalous gauge boson couplings}

\begin{figure}[htbp]
\centering
\includegraphics[width=0.95\columnwidth,height=5.cm]{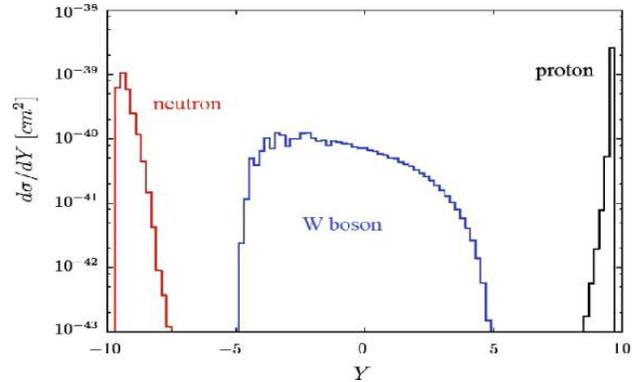}
\caption{Rapidity distribution of single-$W$ photoproduction~\protect\cite{dreyer} 
in $p$-$p$ at $\sqrts$~=~14~TeV -- note that the forward protons (neutrons) fall 
within the RP (ZDC) acceptances.}
\label{fig:gW}
\end{figure}

The study of the couplings of the gauge bosons ($\gamma$, $W$ and $Z$) among
themselves provides the most direct (yet difficult) way to test the gauge structure 
of the electroweak theory. 
Many physics scenarios beyond the SM, with novel interactions and/or particles, 
lead to modifications of the gauge boson self-interaction vertices. 
A process well-suited to testing the triple $WW \gamma$ vertex is 
the photoproduction of single $W$ bosons from a nucleon in ultra-peripheral 
$p$-$p$~\cite{louvain} and $A$-$A$~\cite{dreyer} collisions tagged 
with forward protons or neutrons (Fig.~\ref{fig:gW}). A large cross section of about 
1~pb is expected for large photon-proton c.m. energies, $W_{\gamma\,p}>$~1~TeV. 
In addition, the exclusive two-photon production of $W^+W^-$ pairs probes {\it quartic} 
gauge-boson-couplings. The process has a total cross section of more than 100~fb, and a 
very clear signature. Its cross section is still about 10~fb for $W_{\gamma\,p}>$~1~TeV 
showing sensitivity to physics beyond the SM~\cite{louvain}.

\section{Higgs physics}
\label{sec:higgs}

\subsection{Vector Boson Fusion production}

\begin{figure}[htbp]
\centering
\includegraphics[width=0.95\columnwidth]{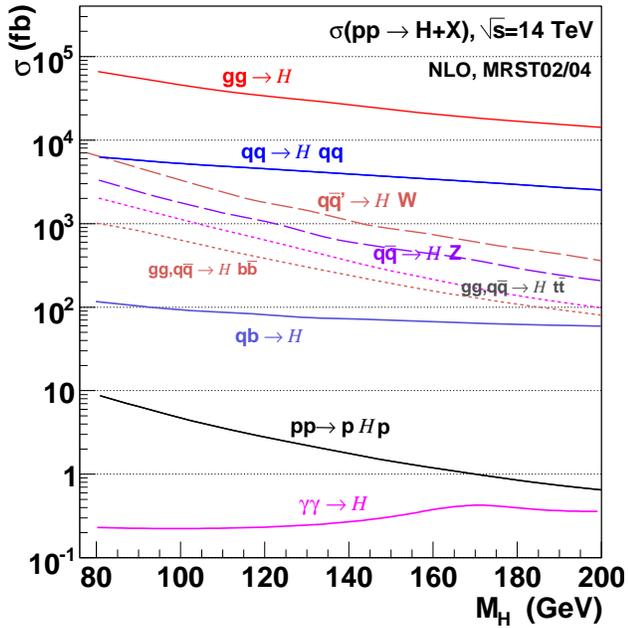}
\vspace{-0.3cm}
\caption{Cross sections for the SM Higgs versus $M_h$ in nine different production channels
at the LHC, from~\protect\cite{spira} ($\sigma_{pHp}$ is from~\cite{fp420}, 
and $\sigma_{\gaga\to H}$ from~\protect\cite{Papageorgiu:1995eg}).}
\label{fig:sigmaHiggs}
\end{figure}

\begin{figure}[htbp]
\centering
\includegraphics[width=0.85\columnwidth,height=6.cm]{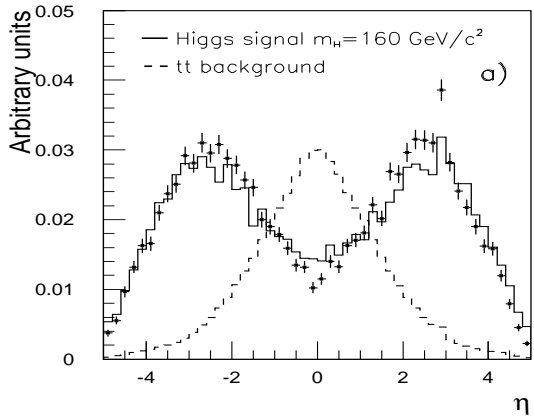}
\vspace{-0.7cm}
\caption{Typical pseudorapidity distribution of the jets in VBF Higgs events 
and in $t\bar{t}$ background events~\protect\cite{Asai:2004ws}.}
\label{fig:VBF}
\end{figure}

The second most important production channel of the SM Higgs boson at the LHC 
(Fig.~\ref{fig:sigmaHiggs}), is the vector-boson-fusion (VBF) process where two $W$ or $Z$, 
radiated off valence quarks, merge to produce a Higgs. The distinctive signature of VBF 
is the presence of two forward-backward jets from the fragmentation of the two 
incoming quarks. The importance of forward calorimetry proves crucial to select VBF events.
The average separation of these jets is $\Delta\eta\approx$~5 and, thus, VBF events 
are usually tagged in ATLAS and CMS by jets detected in the FCal and HF calorimeters. 
Such a distinct event topology reduces efficiently a significant fraction of the Higgs backgrounds  
(e.g. $t \bar t$, $WW$ or $W,Z$+$nj$ which are characterized by jets at {\it central} rapidities, 
see Fig.~\ref{fig:VBF}) making the $qq\,H\rightarrow qq\,WW$ and $qq\,H\rightarrow qq\,\tau\tau$ 
discovery channels at the LHC~\cite{CMS_TDR2,Asai:2004ws}. 

\subsection{Central exclusive production}

Central exclusive production (CEP) is defined as a process of the type 
$p p \rightarrow p \oplus X \oplus p$, where $X$ is a fully measured simple 
state such as $\chi_{c,b}$, jet-jet ($j\,j$), $\gaga$, H, ... and '$\oplus$' represents a 
large rapidity-gap ($\Delta\eta\gtrsim$ 4). Central exclusive Higgs production 
(diagram (d) of Fig.~\ref{fig:feyn_diags}) has attracted significant experimental 
and theoretical interest 
for various important reasons~\cite{DeRoeck:2002hk,fp420}. 
First, to a very good approximation in CEP the primary active di-gluon system obeys a 
$J_z$ = 0, C-even, P-even, selection rule (where $J_z$ is the projection of the total angular 
momentum along the proton beam axis). This selection rule readily permits a 
clean determination of the quantum numbers of any new resonance by measuring the azimuthal 
correlations of the scattered protons. Second, because the process is exclusive, the energy loss 
of the outgoing protons is directly related to the invariant mass of the central system, $M^2 \approx s \xi_1 \xi_2$, 
allowing an excellent mass measurement ($\sigma_{M}\sim 2$~GeV/c$^{2}$) irrespective of the decay 
mode of the central system. Third, thanks to the spin selection rule a large fraction of QCD 
production is suppressed resulting in a very favorable 1:1 signal-to-background. 
The expected SM cross sections are of order 3--10~fb (Fig.~\ref{fig:sigmaHiggs})
although, in certain regions of the minimal supersymmetric extension of the SM (MSSM, at high 
$\tan\beta$ and small $M_A$) with enhanced Higgs coupling to fermions, they can be a factor of 
10--100 larger~\cite{Heinemeyer:2007tu}.\\

\begin{figure}[htbp]
\centering
\includegraphics[width=0.99\columnwidth,height=4.5cm]{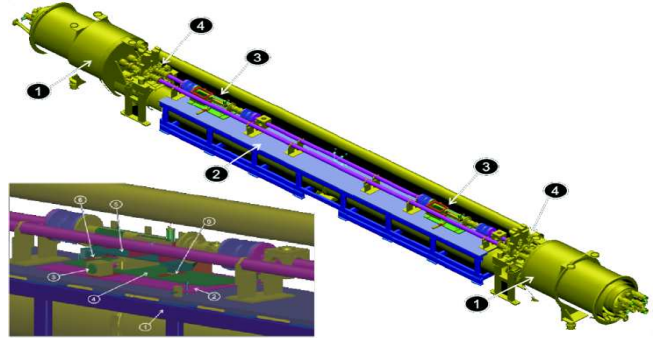}
\caption{Top view of the proposed FP420 system at 420~m from ATLAS and CMS IPs (zoom shows the support 
table with one detector section)~\cite{fp420}.}
\label{fig:fp420}
\end{figure}

\begin{figure}[htbp]
\centering
\includegraphics[width=0.8\columnwidth,height=4.5cm]{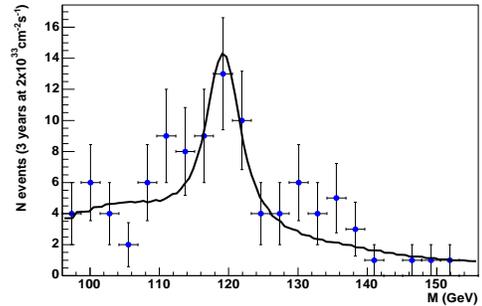}
\caption{Expected mass fit for the MSSM $h\rightarrow b\bar{b}$ decay ($M_h$~=~120~GeV/c$^{2}$), 
measured with FP420 in 60~fb$^{-1}$ integrated luminosity~\cite{fp420,Cox:2007sw}.
(The significance of the fit is $3.5 \sigma$).}
\label{fig:CEP}
\end{figure}

For a Higgs mass close to the LEP limit, $M_H\approx$~120~GeV/c$^2$, the optimal proton tagging 
acceptance is beyond the Roman Pots at 220,240~m.
The proposed FP420 detector system~\cite{fp420} -- a magnetic spectrometer consisting of a moveable 
silicon tracking system and fast \v{C}erenkov detectors located in a 12-m region at about 420~m from 
the ATLAS and CMS IPs (Fig.~\ref{fig:fp420}) --  allows for the detection of both outgoing protons 
scattered by a few hundreds $\mu$rads (i.e. 3 -- 9~mm) relative to the LHC beam line.
A measurement of the protons relative time of arrival in the 10~ps range is required 
for matching them with a central vertex within $\sim$2 mm, which will enable the 
rejection of a large fraction of the pile-up background at high-luminosities. Under such
circumstances, the Higgs boson line-shape can be reconstructed in the (difficult) $b\bar{b}$
channel with a $3\sigma$ or better significance with an integrated luminosity of 60~fb$^{-1}$
(Fig.~\ref{fig:CEP}).

\section*{Summary}

Many interesting scattering processes at the LHC -- mediated by colorless exchanges (photons, 
heavy gauge-bosons, Pomerons, di-gluons in a color-single-state) -- are open to study 
at TeV energies for the first time, thanks to detector instrumentation 
at low angles with respect to the beam. We have reviewed the near-beam instrumentation 
capabilities (at pseudo-rapidities $|\eta| >$~3) and the associated ``forward'' physics programme 
of the various LHC experiments in various sectors and extensions of the Standard Model.
The varied and unique set of measurements accessible in the QCD (diffractive, low-$x$, 
hadronic MCs for cosmic-rays), electroweak (exclusive dileptons, gauge couplings), 
Higgs (vector-boson-fusion and exclusive production) and beyond the SM (anomalous 
couplings, exclusive MSSM Higgs) sectors, provides a powerful and complementary
way to explore the particles and interactions of nature at energies never reached before.

\begin{acknowledgments}

I would like to thank the organisers of LAWHEP'07 -- in particular Beatriz Gay-Ducati,
Magno Machado and Carlos A. Garcia-Canal -- for their kind invitation to an stimulating 
physics school. Special thanks due to Michele Arneodo and Albert de Roeck for valuable discussions.
Work supported by the 6th EU Framework Programme contract MEIF-CT-2005-025073.
\end{acknowledgments}

\appendix

\def\IJMPA{{Int. J. Mod. Phys.}~{\bf A}}
\def\EPJ{{Eur. Phys. J.}~{\bf C}}
\def\JPG{{J. Phys.}~{\bf G}}
\def\JHEP{{J. High Energy Phys.}~}
\def\NCA{Nuovo Cimento~}
\def\NIM{Nucl. Instrum. Methods~}
\def\NIMA{{Nucl. Instrum. Methods}~{\bf A}}
\def\NPA{{Nucl. Phys.}~{\bf A}}
\def\NPB{{Nucl. Phys.}~{\bf B}}
\def\PLB{{Phys. Lett.}~{\bf B}}
\def\PLC{Phys. Repts.\ }
\def\PRL{Phys. Rev. Lett.\ }
\def\PRD{{Phys. Rev.}~{\bf D}}
\def\PRC{{Phys. Rev.}~{\bf C}}
\def\ZPC{{Z. Phys.}~{\bf C}}

\begin{footnotesize}

\end{footnotesize}

\end{document}